\documentclass[twocolumn, tighten]{aastex63}

\received{July 23, 2021}
\revised{September 29, 2021}
\accepted{October 8, 2021}
\submitjournal{ApJL}

\graphicspath{{./}{./}}

\begin{document}

\title{\sc Milky Way-Like Gas Excitation in an Ultrabright Submillimeter Galaxy at $z=1.6$}

\correspondingauthor{Nikolaus Sulzenauer}
\email{nsulzenauer@mpifr-bonn.mpg.de}

\author[0000-0002-3187-1648]{N. Sulzenauer}
\affiliation{University of Vienna, Department of Astrophysics, T\"urkenschanzstrasse 17, A-1180, Vienna, Austria}
\affiliation{Instituto de Astrofísica de Canarias (IAC), E-38205 La Laguna, Tenerife, Spain}
\affiliation{Universidad de La Laguna, Dpto. Astrofísica, E-38206 La Laguna, Tenerife, Spain}
\affiliation{Max-Planck-Institut für Radioastronomie, Auf dem Hügel 69, 53121 Bonn, Germany}

\author[0000-0001-7147-3575]{H. Dannerbauer}
\affiliation{Instituto de Astrofísica de Canarias (IAC), E-38205 La Laguna, Tenerife, Spain}
\affiliation{Universidad de La Laguna, Dpto. Astrofísica, E-38206 La Laguna, Tenerife, Spain}

\author[0000-0003-0748-4768]{A. Díaz-Sánchez}
\affiliation{Departamento Física Aplicada, Universidad Politécnica de Cartagena, Campus Muralla del Mar, E-30202 Cartagena, Murcia, Spain}

\author[0000-0003-2856-1080]{B. Ziegler}
\affiliation{University of Vienna, Department of Astrophysics, T\"urkenschanzstrasse 17, A-1180, Vienna, Austria}

\author{S. Iglesias-Groth}
\affiliation{Instituto de Astrofísica de Canarias (IAC), E-38205 La Laguna, Tenerife, Spain}
\affiliation{Universidad de La Laguna, Dpto. Astrofísica, E-38206 La Laguna, Tenerife, Spain}

\author[0000-0003-3767-7085]{R. Rebolo}
\affiliation{Instituto de Astrofísica de Canarias (IAC), E-38205 La Laguna, Tenerife, Spain}
\affiliation{Universidad de La Laguna, Dpto. Astrofísica, E-38206 La Laguna, Tenerife, Spain}
\affiliation{Consejo Superior de Investigaciones Cientficas, E-28006 Madrid, Spain}



\begin{abstract}
Based on observations with the IRAM 30~m and Yebes 40~m telescopes, we report evidence of the detection of Milky Way-like, low-excitation molecular gas, up to the transition CO($J=5-4$), in a distant, dusty star-forming galaxy at $z_{CO}=1.60454$. \textit{WISE} J122651.0+214958.8 (alias SDSSJ1226, the \emph{Cosmic Seahorse}), is strongly lensed by a foreground galaxy cluster at $z=0.44$ with a source magnification of $\mu=9.5\pm0.7$. This galaxy was selected by cross-correlating near-to-mid infrared colours within the full-sky \textit{AllWISE} survey, originally aiming to discover rare analogs of the archetypical strongly lensed submillimeter galaxy SMM J2135-0102, the Cosmic Eyelash.  We derive an apparent (i.e. not corrected for lensing magnification) rest-frame 8-1000 $\mu$m infrared luminosity of $\mu L_\mathrm{IR}=1.66^{+0.04}_{-0.04}\times 10^{13}$ L$_\sun$ and apparent star-formation rate $\mu\mathrm{SFR}_\mathrm{IR}=2960\pm70$ M$_\sun$ yr$^{-1}$. SDSSJ1226 is ultra-bright at $S_{350\mu m}\simeq170$ mJy and shows similarly bright low-$J$ CO line intensities as SMM J2135-0102, however, with exceptionally small CO($J=5-4$) intensity. We consider different scenarios to reconcile our observations with typical findings of high-$z$ starbursts, and speculate about the presence of a previously unseen star-formation mechanism in cosmic noon submillimeter galaxies. In conclusion, the remarkable low line luminosity ratio $r_{5,2}=0.11\pm0.02$ is best explained by an extended, main-sequence star-formation mode -- representing a missing link between starbursts to low-luminosity systems during the epoch of peak star-formation history. 
\end{abstract}



\setcounter{footnote}{5}

\section{Introduction} \label{sec:intro}
To get insight into the formation and evolution of massive galaxies at the peak epoch of star formation and black hole activity it is indispensable to study dusty starbursts, so-called submillimeter galaxies \citep[SMGs; see for a review][]{Casey2014}. These dusty star-forming galaxies at a median redshift of $z=2.3$ \citep{Chapman2005} are rich in molecular gas \citep{Tacconi2008}, the fuel for star formation. The brightest of these systems are either intrinsically luminous with extreme star-formation rates in excess of several hundred to thousand solar masses per year and/or galaxies that are strongly gravitational lensed by chance alignments with foreground galaxies or galaxy clusters \citep{Negrello2010}. The boosted apparent (sub)millimeter flux of this rare population of low number density $N(S_{500\mu m}>100\;\mathrm{mJy})=0.2$ deg$^{-2}$ and otherwise optically faint or even undetected sources \citep{Dannerbauer2002}, is efficiently identified in wide \textit{Herschel} or \textit{Planck} space missions, and South Pole Telescope surveys \citep[see e.g.][]{Negrello2010,Vieira2013,Canameras2015}.

A remarkable example of a strongly lensed SMG is the serendipitously discovered ultra-bright SMM J2135-0102 at $z=2.326$, the so-called "Cosmic Eyelash" \citep{Swinbank2011}. Resolved observations of cold gas tracers such as molecular carbon monoxide (CO) were subsequently utilised in lensed galaxies to test theories of star-formation and the conditions of the cold interstellar medium at early cosmic epochs. The assumption that all SMGs are simply scaled-up versions of local Universe ULIRGs -- mostly gas rich major mergers \citep{Papadopoulos2012} that form stars in compact, nuclear disks -- is increasingly challenged as high-redshift molecular clouds in star-forming galaxies are observed to inherit their properties from extended, fragmented gas disks, forming massive clumps at a scale of $\lesssim100$ pc \citep[see e.g.][]{Daddi2015,Dessauges-Zavadsky2019}. Yet, \citet{Ivison2020} reported that the disk of the Cosmic Eyelash is probably smoothly distributed, at least down $\sim$80 pc-scale, reinforcing circumstantial evidence for the heterogeneous population of SMGs, differentiated by their modes of star-formation. 

In this regards, gravitational lensing promises the detection of low-luminosity systems. Compared to classical SMGs, these systems are forming stars within extended gas disks at an order of magnitude lower efficiency \citep{Tacconi2010}. Characterised by low CO excitation profiles (CO SLEDs), as reported by \citet{Dannerbauer2009} in a sample of $z\approx1.5$ colour-selected galaxies \citep{Daddi2015} and in resolved studies of CO gas in the lensed disk galaxy Cosmic Snake at $z=1.036$ \citep{Dessauges-Zavadsky2019}, less efficient, slow-mode star-formation should be present in low-luminosity, normal galaxies. However, the number of CO detections of these main-sequence galaxies at high-redshift is still relatively small \citep{Valentino2020} and strong evidence for true Milky Way-like cold gas excitation up to high CO-transitions is still lacking or at least inconclusive for this galaxy population.

In this Letter, we present the unusually low, unambiguous Milky Way-like CO excitation profile of SDSSJ1226, an ultra-bright submm galaxy at $z=1.60454$, strongly lensed by a galaxy cluster at $z=0.44$. In section \ref{sec:sample}, we present the sample and in section \ref{sec:observations} the observations. Results are shown in section \ref{sec:results} and we conclude this manuscript with the discussion (section \ref{sec:discussion}). A more complete analysis of the subsample of \textit{Herschel}-detected, NIR/MIR SMG candidates (including this source) will be presented in Sulzenauer et al. (in prep.). In this work we focus on the CO SLED of SDSSJ1226. We adopt a flat $\Lambda$CDM cosmology using parameters from \citet{Planck2016} with $H_0=67.8\pm0.9$ km s$^{-1}$ Mpc$^{-1}$, $\Omega_m=0.308\pm0.012$, and $\Omega_\Lambda=1-\Omega_m$.

\section{Sample selection}\label{sec:sample}
\setcounter{footnote}{5}

In order to search for bright SMGs with similar SED characteristics as the archetypical bright, strongly lensed SMG SMM J2135-0102, \emph{the Cosmic Eyelash}, \citet{Iglesias-Groth2017} developed a NIR/MIR colour selection technique to cross-match MIR sources over the full-sky by correlating \textit{AllWISE}\footnote{\url{http://wise2.ipac.caltech.edu/docs/release/allwise/}} catalog \citep{Wrigth2010} (NIR to MIR colours) from the \textit{Wide-field Infrared Survey Explorer}. Galaxies that verify the NIR/MIR colour criterion from \citet{Iglesias-Groth2017} 
\begin{eqnarray*}
J-K_{s}>2.0\;\wedge\;K_{s}-W1>1.4\;\wedge\\
W1-W2>0.8\;\wedge\;W2-W3<2.4\;\wedge\;W3-W4>3.5
\end{eqnarray*}
(in AB-magitudes)
are pre-selected as "Cosmic Eyelash analog" candidates at $z\approx2$. Subsequent source matching with additional NIR, FIR, and submillimeter data has proven to efficiently select bright SMGs: \citet{Diaz-Sanchez2017} identified the extremely bright SMG WISE J132934.13+234327.3, the so-called Cosmic Eyebrow, by cross-matching between NIR/MIR selected candidates,  \textit{Planck}\footnote{\url{http://pla.esac.esa.int/pla/home}} full-sky point source catalog, and JCMT/SCUBA-2\footnote{\url{http://www.cadc-ccda.hia-iha.nrc-cnrc.gc.ca/en/}} data. Spectroscopic follow-up observations with the IRAM 30~m telescope and IRAM NOEMA detected the brightest CO($J=3-2$) emission ever for SMGs \citep{Dannerbauer2019}. As part of the same sample, the strongly lensed source GAL-CLUS-022058s was followed up with APEX/nFLASH230 spectroscopy confirming extremely bright CO($J=5-4$) emission \citep{Diaz-Sanchez2021}. 

This method was further modified by correlating \textit{Herschel}/SPIRE\footnote{\url{http://archives.esac.esa.int/hsa/whsa/}} sources with NIR/MIR candidates, that do not necessarily match all five colour criteria, but are in close distance to 28 strong lensing clusters from the Sloan Giant Arcs Survey \citep{Oguri2012}.
Due to the availability of archival \textit{HST} optical data for all of the clusters, the giant arcs were further matched with bright, co-spatial \textit{AllWISE} candidates within 2$\arcsec$. We have found eight candidates, with $1.5<z_\mathrm{phot}<2.5$, two of them have high apparent infrared luminosities $\mu L_{IR} > 10^{13}$ L$_\odot$ and flux $S_{350\mu m} > 100$ mJy whereas \textit{AllWISE} J122651.04+214958.8 is the brightest of the sample with FIR flux $S_{350\mu m}=170$ mJy. Although, it does neither fulfil the NIR-color criteria from \citet{Iglesias-Groth2017} nor the one from \citet{Diaz-Sanchez2017}, which are based on the Cosmic Eyelash SED, the \textit{Herschel} FIR flux does agree very well with the behavior of SMG SEDs. This can be explained by large galaxy-to-galaxy variation in the optical-NIR domain due to the combination of dust geometry, star-formation history, and extinction \citep[see e.g.][]{DaCunha2015}. The main discrepancy between the formal NIR/MIR colour demarcations and the colour indices found for galaxy SDSSJ1226 are shallower indices $W1-W2\approx0.3$ instead of $>0.8$ and $W3-W4\approx2.2$ whereas $W4$ is only detected as an upper limit implying that the true index might be even more distant from the demarcation index $W3-W4>3.5$. Furthermore, $W2-W3\approx-0.2$, although within the criterion definition,  is especially low compared to the reported sample values of \citet{Iglesias-Groth2017}. Nevertheless, SDSSJ1226 is consistently brighter than the Cosmic Eyelash  by a factor of $\sim$2 in all \textit{WISE} bands, but shows lower flux towards far-infrared wavelengths; see Table \ref{tab:photometry} for additional photometric catalog data from VizieR at CDS\footnote{\url{http://vizier.u-strasbg.fr/viz-bin/VizieR}}.

\begin{deluxetable*}{ccccl}
\tablenum{1}
\tabletypesize{\footnotesize}
\tablecaption{Archival photometry associated with source SDSSJ1226.\label{tab:photometry}}
\tablehead{
\colhead{Wavelength [$\mu$m]} & \colhead{Flux Density$^\mathrm{a}$ [mJy]} & \colhead{AB-Mag.$^\mathrm{a,d}$ [mag]} & \colhead{Beam Size [\arcsec]} & \colhead{Observatory/Instrument} 
}
\decimalcolnumbers
\startdata
0.577 & 0.0007$\pm$0.0001 & 24.3$\pm$0.15    & $\sim$0.1 & \textit{HST}/ACS F606W \\
0.797 & 0.0022$\pm$0.0002 & 23.0$\pm$0.1    & $\sim$0.1 & \textit{HST}/ACS F814W \\
1.248 & 0.020$\pm$0.002 & 20.6$\pm$0.11    & $<$1.2 & UKIDSS $J$ \\
1.635 & 0.034$\pm$0.003 & 20.1$\pm$0.1    & $<$1.2 & UKIDSS $H$ \\
2.201 & 0.13$\pm$0.03 & 18.6$\pm$0.26    & $<$1.2 & UKIDSS $K_s$ \\
3.4 & 0.496$\pm$0.019$^\mathrm{b}$ & 17.2$\pm$0.04 & 6.1 & \textit{WISE} W1 \\
4.6 & 0.653$\pm$0.025$^\mathrm{b}$ & 16.9$\pm$0.04 & 6.4 & \textit{WISE} W2 \\
11.6 & 0.525$\pm$0.13$^\mathrm{b}$ & 17.1$\pm$0.27 & 6.5 & \textit{WISE} W3 \\
22.1 & $\sim$3.8$^\mathrm{b,c}$ & $\sim$14.9$^\mathrm{c}$    & 12.0 & \textit{WISE} W4 \\
160 & 87$\pm$31 & 11.5$\pm$0.4    & 11.6 & \textit{Herschel}/PACS \\
250 & 172$\pm$9 & 10.8$\pm$0.06    & 18.5 & \textit{Herschel}/SPIRE \\
350 & 170$\pm$13 & 10.8$\pm$0.08   & 25.3 & \textit{Herschel}/SPIRE \\
500 & 115$\pm$12 & 11.2$\pm$0.1    & 36.9 & \textit{Herschel}/SPIRE \\
\enddata
\tablecomments{
$^\mathrm{a}$Uncorrected for source magnification. $^\mathrm{b}$Flux densities are calculated using the prescription from \citet{Wrigth2010}. $^\mathrm{c}$Upper limit. $^\mathrm{d}$AB-magnitudes are calculated using $\mathrm{AB}=-2.5\times\log_{10}{S_\nu/\mathrm{[Jy]}}+8.9$ and uncertainties are from a linear approximation. 
}
\end{deluxetable*}

\begin{figure}
\epsscale{1.15}
\plotone{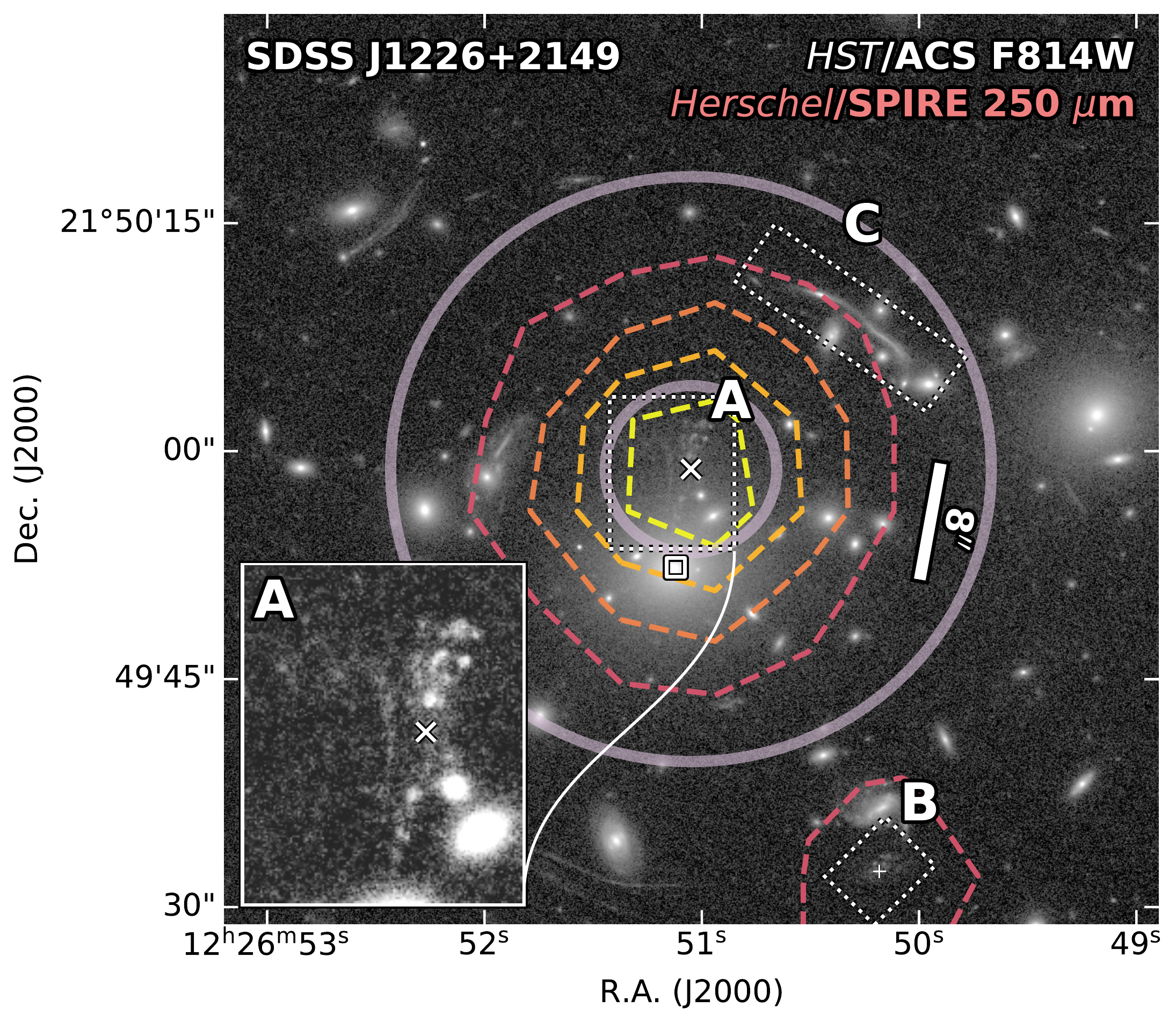}
\caption{Background image centred on the \emph{Herschel}/SPIRE flux centroid of source SDSSJ1226-A (\emph{Cosmic Seahorse}) marked by the white cross. \textit{HST}/ACS F814W data is used to show the presumed optical counterpart (inset with stellar foreground emission gradient masked out) of the FIR bright source, visible as a strongly lensed radial arcs projected north to the BCG (square) of a strongly lensing galaxy cluster at $z=0.435\pm0.009$ \citep{Bayliss2011,Oguri2012}. Boxes denote main image A, possible counter-image B (white plus), and arc SDSSJ1226-C that belongs to a different galaxy at similar redshift, $z=1.605$, as SDSSJ1226-A \citep{Oguri2012}. \emph{Herschel}/SPIRE 250 $\mu$m contour levels at $[2,3,4,5]\sigma$ confidence intervals above the local background noise level of $\sigma\approx3.4$ mJy/beam trace the bright FIR source's location. The IRAM 30~m telescope HPBW at 221 GHz ($\approx 11.6\arcsec$), corresponding to CO($J=5-4$), is marked by the inner lilac ring, while the outer circle shows the lowest resolution single-dish HPBW, CO($J=1-0$) from Yebes 40~m telescope, at $\approx 39.1\arcsec$, all beam sizes fall in between these two values.}
\label{fig:backgroundimage}
\end{figure}

On the basis of its position, colour, and brightness, we confidently identify the supposed optical counterpart of the \textit{Herschel}/SPIRE source in sufficiently resolved \textit{HST}/ACS images\footnote{\url{http://archive.stsci.edu/}}. The corresponding giant radial arc is located $\sim15\arcsec$ north to the brightest cluster galaxy (BCG) in the galaxy cluster SDSS J1226+2149 at redshift $z=0.435\pm 0.009$ \citep{Bayliss2011}. Therefore, we postulate that the strongly lensed galaxy at coordinates $\alpha=12^\mathrm{h}26^\mathrm{m}51\fs 05$ and $\delta=+21\degr 49\arcmin 58\farcs80$ (J2000) is the source of bright NIR--FIR emission. No optical/NIR spectroscopic redshift of this source are published in the literature. Fig. \ref{fig:backgroundimage} shows the \textit{HST}/ACS background image at the position of the bright \textit{Herschel}/SPIRE source. 
Due to its extraordinary brightness and distinctive morphology, we take the liberty to designate this lensed galaxy as the \emph{Cosmic Seahorse}.

\section{Observations and Data Reduction} \label{sec:observations}

\subsection{Observations}
\subsubsection{IRAM 30~m telescope}
We observed the source with the IRAM 30m telescope heterodyne millimeter receiver EMIR \citep{Carter2012} by employing the \emph{spectroscopic blind line search technique}.
Starting with the photometric redshift estimate of $z\sim2$, the CO($J=2-1$) or CO($J=3-2$) emission line is expected to lie within the 3 mm atmospheric window, accessible by the EMIR E0 frontend ($73-117$ GHz). After detecting a strong line at $\approx 88.5$~GHz and assuming the redshifted CO($J=2-1$) transition, we  switched the frontend to E1 ($125-184$ GHz). A second signal was successfully identified in the 2 mm atmospheric window at $\approx 132.8$~GHz, the CO($J=3-2$) line. 
Based on this improved redshift estimate, CO($J=$5-4) emission was then observed with E2 ($202-274$ GHz) at 221.3~GHz. The observations were conducted in position switching mode. We utilised the FTS200 spectrograph as the backend at a frequency resolution of $\sim$200 kHz. The first week of observations (project; 086-18, PI: H. Dannerbauer) was regularly scheduled in visitor mode. Between May and November of 2018, the project entered the observing pool. 
Over six days of observations, a total of 11.0 hours effective on-source integration time was acquired for the presented multi-$J$ CO line measurements.
Overall, weather conditions were acceptable with median optical depth of $\tau_{255\mathrm{ GHz}}\approx0.5$. 
Notably, however, receiver interference caused baseline ripples in the spectra that needed to be carefully removed.

\subsubsection{Yebes 40~m telescope}

Our observations of the redshifted CO($J=1-0$) line (44.3~GHz)  were performed with the new Nanocosmos Q-band receiver, operating between 31.3-50.6 GHz \citep{Tercero2020}, installed at the Yebes 40~m telescope (project; 20B013, PI: H. Dannerbauer). This front-end consists of two high electron mobility transistor (HEMT) cold amplifiers that cover horizontal and vertical polarisation. The signal is obtained using fast Fourier transform spectrometers, covering eight sub-bands of 2.5 GHz bandwidth and continuous 38 kHz resolution. 
Over twelve days of observations between October and November 2020, we collected scans of our source with effective on-source time of 37.2 hours with better than average optical depth of $\tau_{255\mathrm{ GHz}}=0.080$. We report the most distant detection of molecular gas at $z=1.60454$ with the Yebes 40~m telescopes \citep[see e.g.][]{Tercero2020a}.

\subsection{Data reduction}
For on-site data reduction, the software GILDAS\footnote{\url{https://www.iram.fr/IRAMFR/GILDAS}} with package CLASS was used \citep{Pety2005}. To identify spectral lines at the anticipated noise threshold of $T_A^\ast=0.25$ mK, the spectra are binned to 500 km s$^{-1}$ and typical observing time per setting were four hours in total. Significant line signals are masked for manual baseline removal with a polynomial function of order $n=3$ within observed-frame frequency window of $\Delta\nu\approx1.5$ GHz centred on the signal. After gathering all observations from the pool, individual scans were dropped in order to maximise the signal-to-noise content per spectral bin. For this reason, we developed a simple CLASS script \verb|rmse_selection_function| that is publicly available on GitHub\footnote{\url{https://github.com/NiSZR/rmse_selection_function}}. Up to a specific, non-parametric rms threshold, it filters all scans within the 4 GHz sideband structure of the EMIR FTS200 data by their ranked baseline noise contribution. Adjusting to the background noise level, the frequency channels are rebinned between 13 and 33 km s$^{-1}$ (rest-frame) corresponding to 99.7 \% confidence level of the line intensity per bin at the position of the spectral line. To convert the corrected main beam brightness temperature to flux density, we fitted a parabola to the values from \citet{Velilla2017} with sensitivities of approximately 7.56 Jy K$^{-1}$, 6.31 Jy K$^{-1}$, and 6.00 Jy K$^{-1}$ in the 1 mm, 2 mm, and 3 mm atmospheric windows. Yebes 40~m telescope 44.3 GHz observations are converted using the factor 4.82 Jy K$^{-1}$. The total uncertainty of the flux calibration is assumed to be less than 10 \%. Final data visualisation and flux calculation is performed in Python with modules numpy \citep{Virtanen2019} and astropy \citep{Astropy2013}.

\begin{figure*}
\epsscale{1.17}
\centering
\includegraphics[height=0.18\textheight]{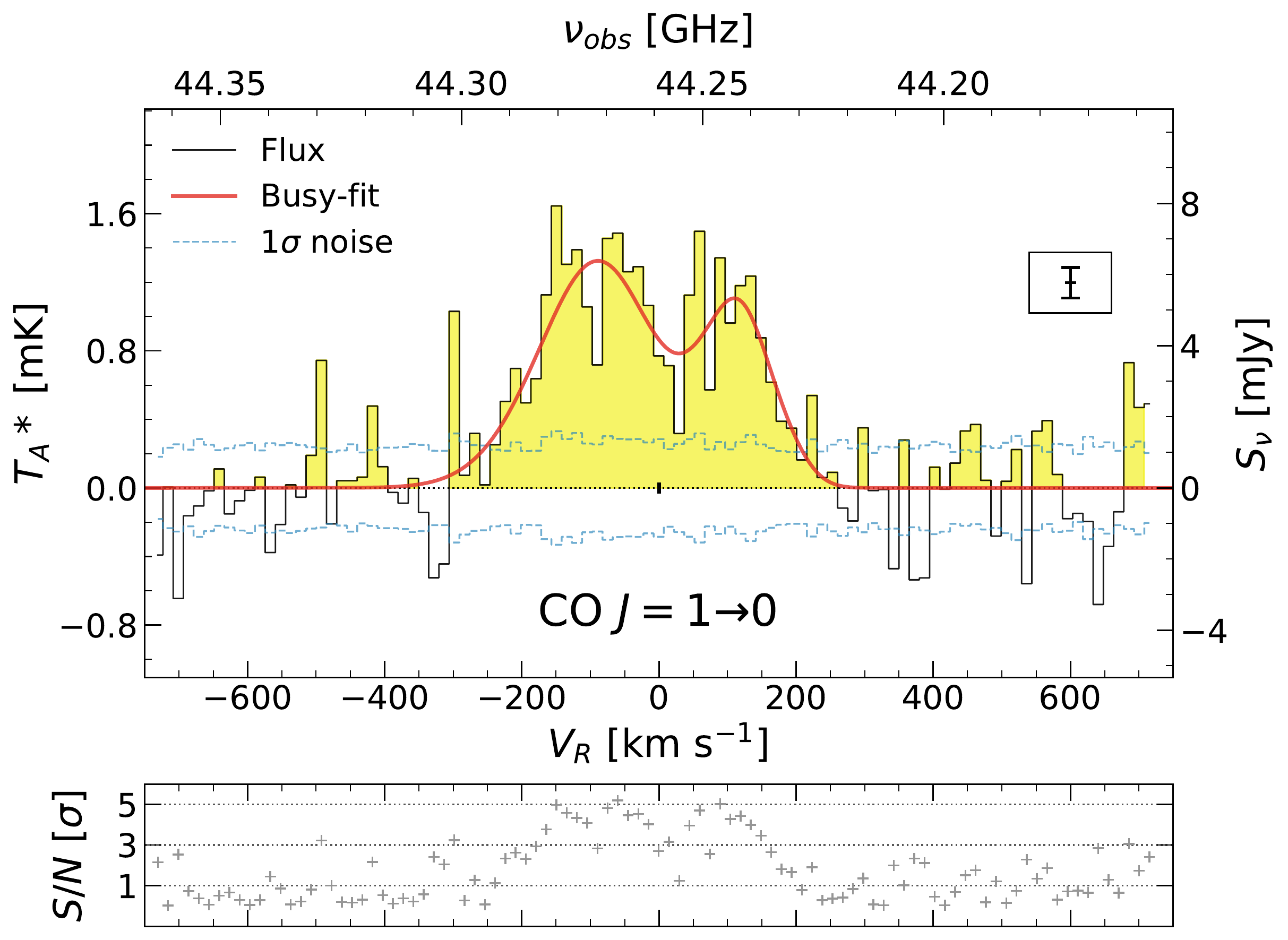}
\includegraphics[height=0.18\textheight]{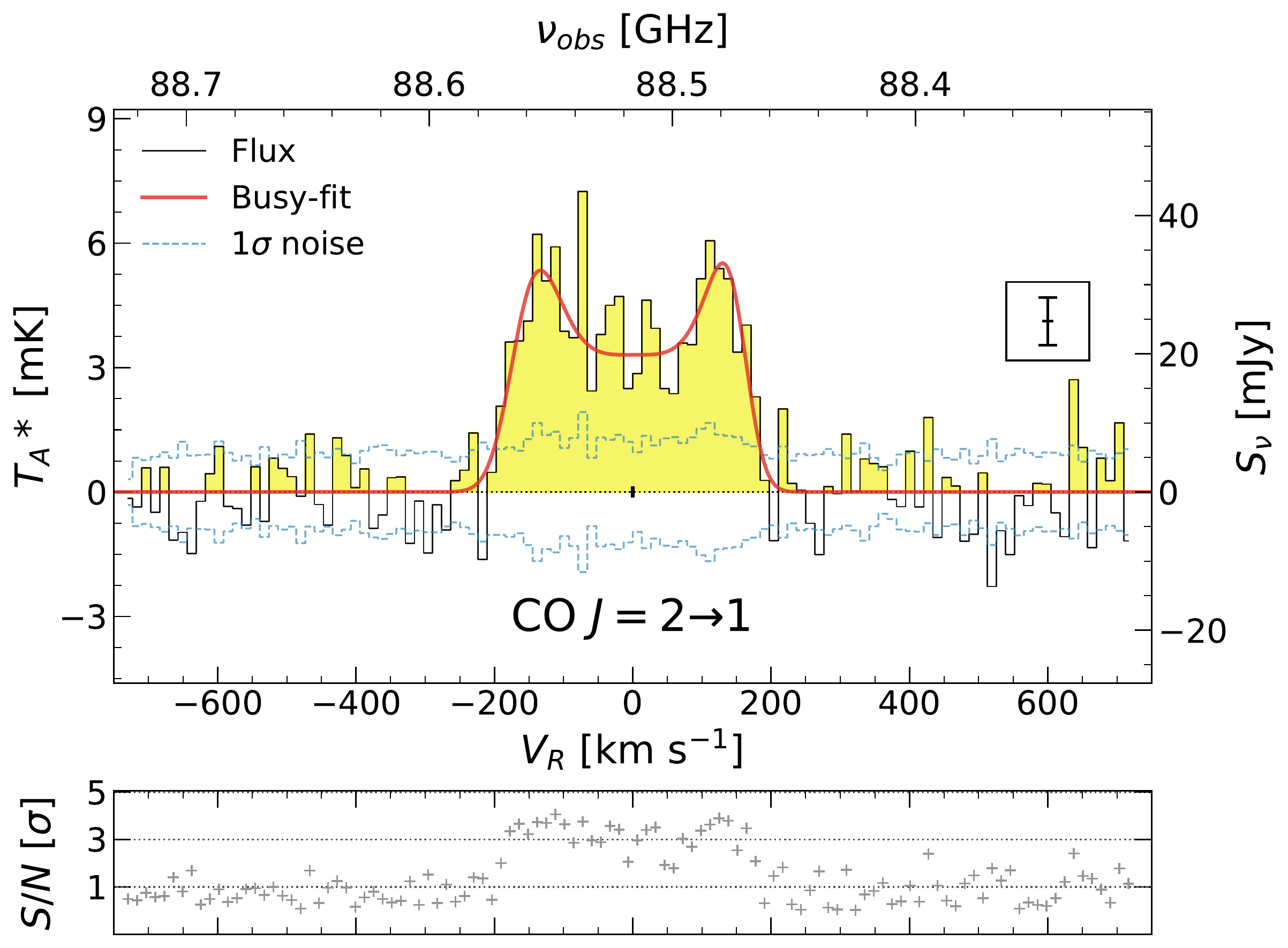}
\includegraphics[height=0.18\textheight]{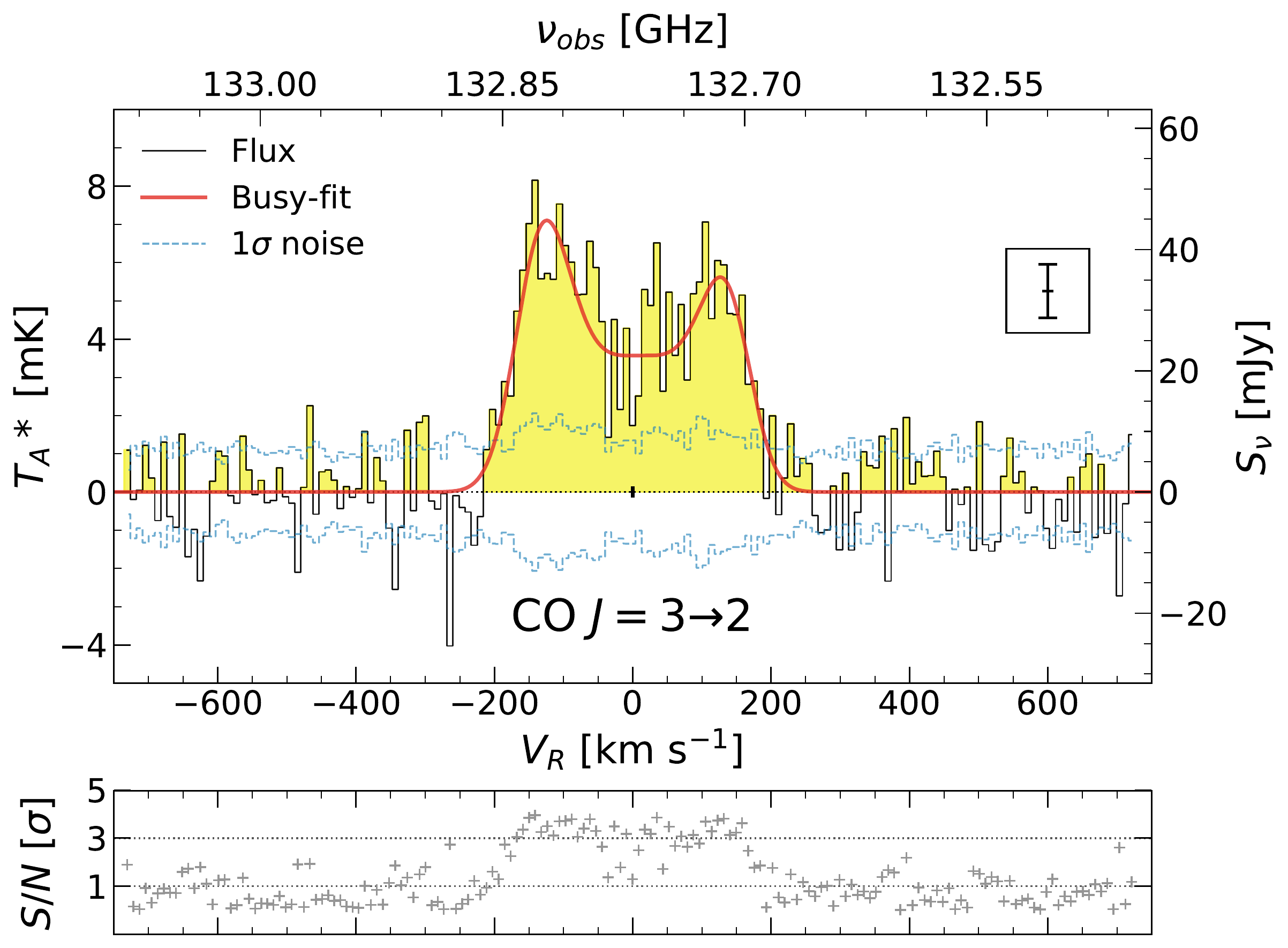}\\
\includegraphics[height=0.18\textheight]{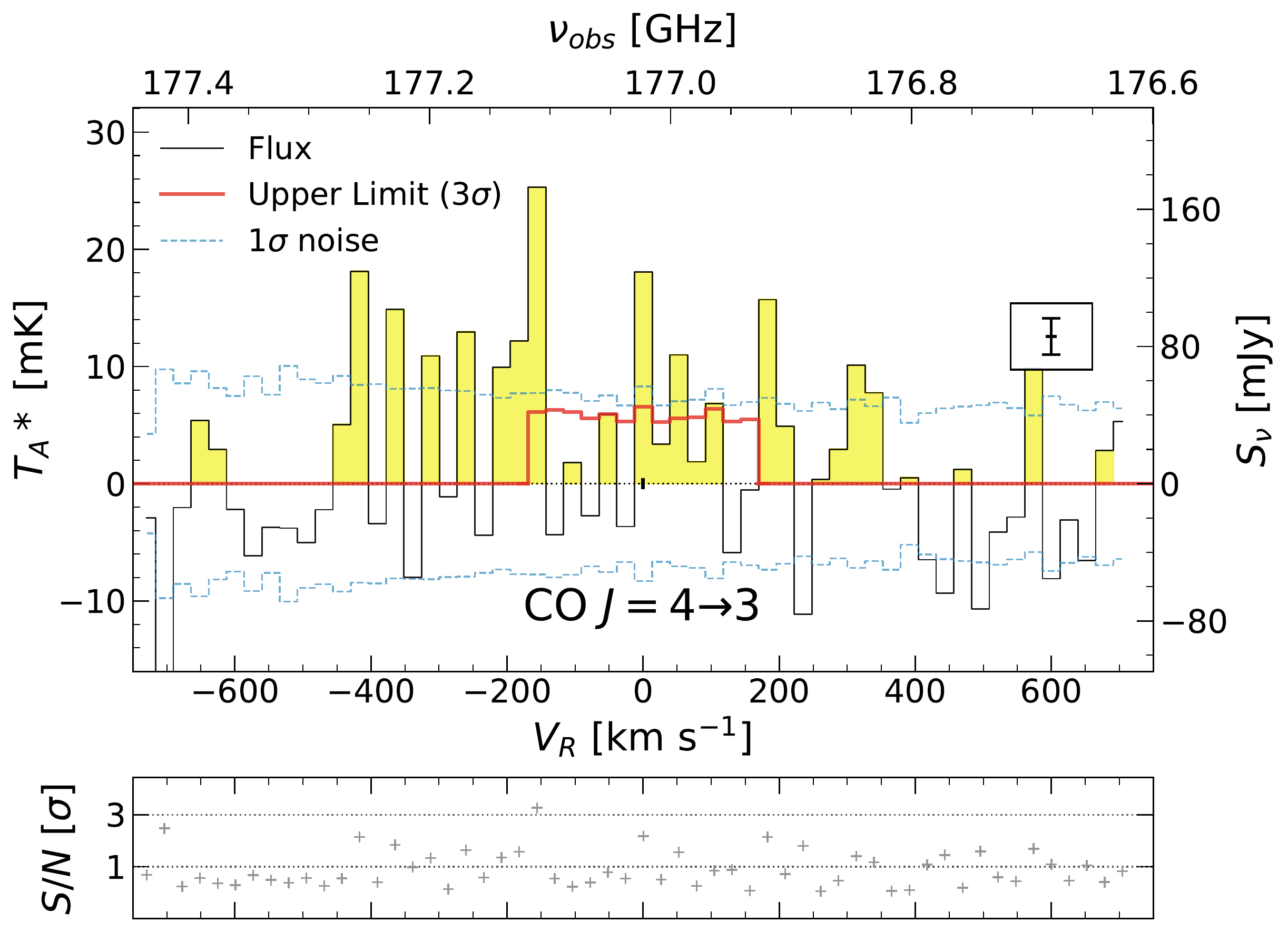}
\includegraphics[height=0.18\textheight]{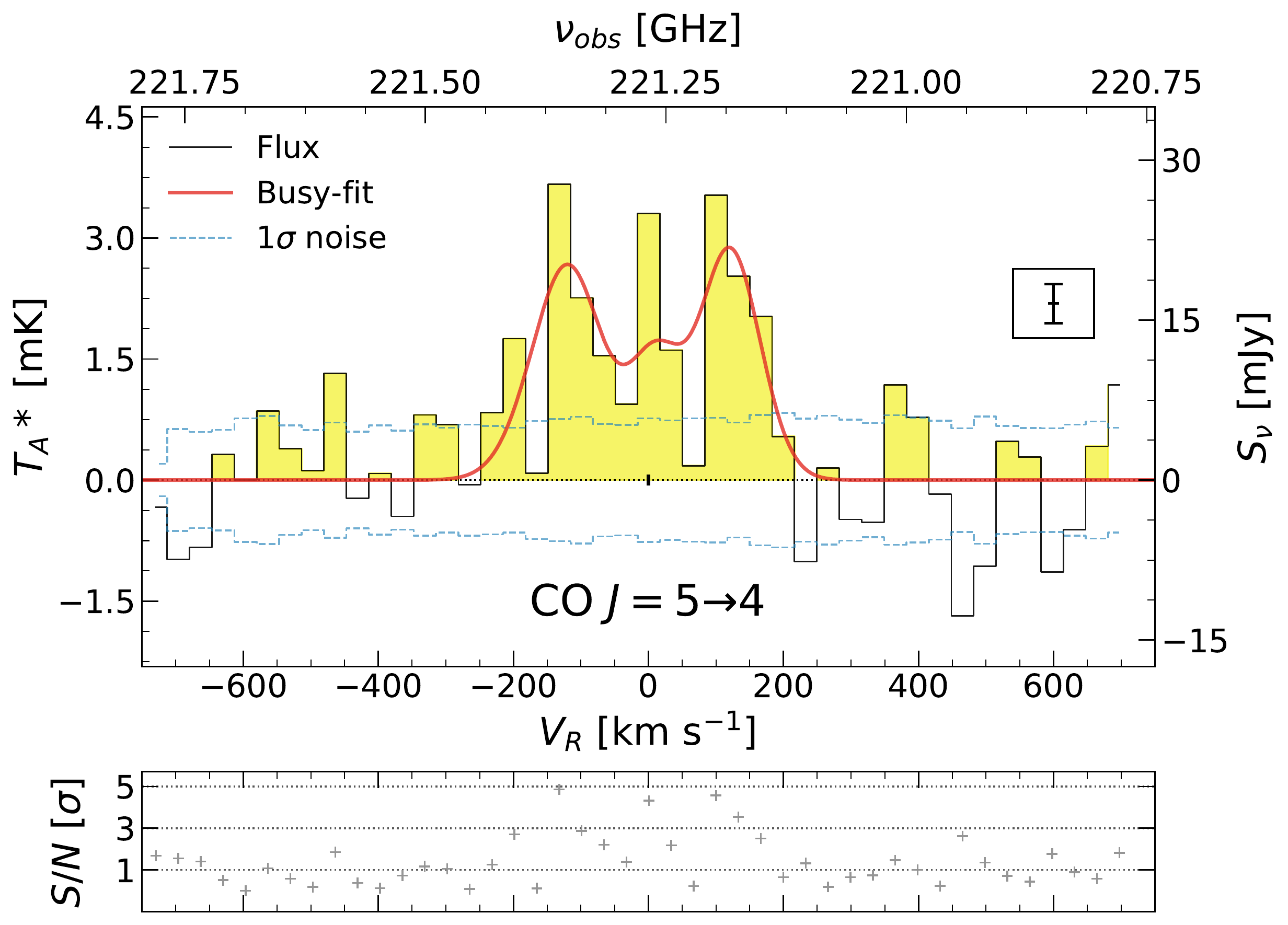}

\caption{CO line spectra of transitions CO($J=1-0$) from  Yebes 40~m radio telescope, and CO($J=2-1$), CO($J=3-2$), CO($J=4-3$), and CO($J=5-4$) from IRAM 30~m telescope in units of Kelvin and milli-Jansky. Boxed error bars give $\pm 2\sigma$ calibration uncertainties. The lower panels show the S/N-values per binned channel in units of background noise rms (blue dashed spectrum). Rms noise levels per channel bins are $[1.20,6.2,7.9,50.5,5.3]$ mJy, respectively. Red curves are flux models fits that are used to calculate the velocity integrated fluxes. Flux of CO($J=4-3$) is regarded as a $3\sigma$ upper-limit measurement.} 
\label{fig:spectra}
\end{figure*}

\section{Results}\label{sec:results}
\subsection{CO SLED}\label{sec:spec}
\begin{figure}
\epsscale{1.17}
\plotone{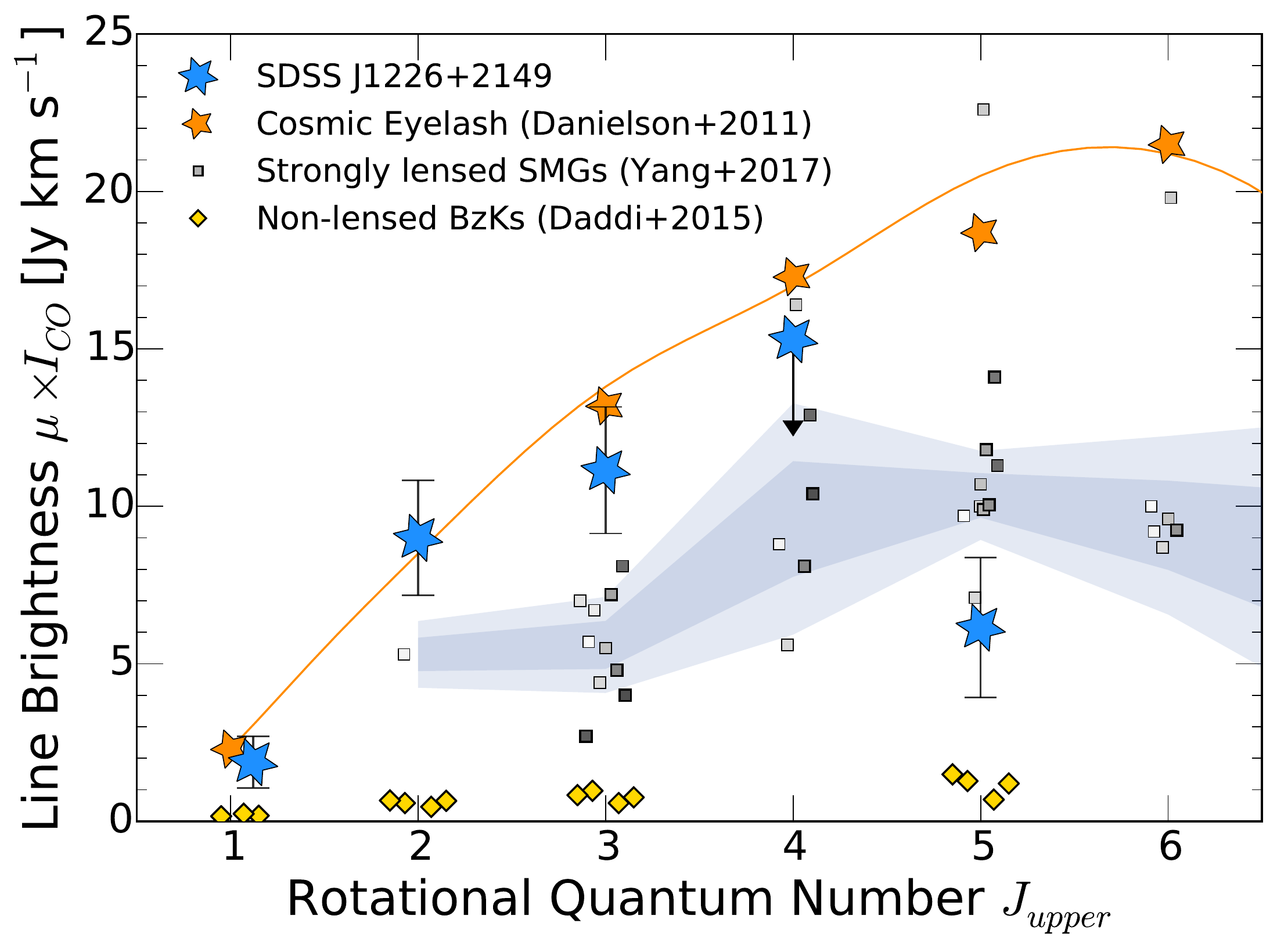}
\caption{Velocity integrated line intensities $I_{CO}\equiv\int S_{CO}\mathrm{d}v$ comparison of cosmic noon galaxies. IRAM 30~m and Yebes 40~m telescopes measurements of SDSSJ1226 (blue stars) are compared to the results of SMM J2135-0102 magnified by $\mu=32$ (orange stars); the orange curve corresponds to the LVG model presented in \citet{Danielson2011}. Error bars are shown at 3$\sigma$ confidence interval. Velocity integrated Intensity for CO($J=4-3$) is given as a $3\sigma$ upper limit. At $J_{upper}\leq3$, SDSSJ1226 is (nearly) as bright as the Cosmic Eyelash and substantially brighter than the \textit{Herschel}-selected, lensed SMG sample from \citet{Yang2017} (grey squares). Shaded regions indicate the 1$\sigma$ and 2$\sigma$ confidence level of the bootstrapped intensity distribution of the aforementioned SMG sample. Yellow diamonds are \citet{Daddi2015} non-lensed, normal star-forming galaxies.}
\label{fig:ICO}
\end{figure}

Fig. \ref{fig:spectra} shows the flux densities for the detected CO spectra of SDSSJ1226 and Fig. \ref{fig:ICO} gives a comparison between the apparent CO line intensities with error bars at a 3$\sigma$ confidence level. We were able to blindly identify four bright emission lines with intensity $I_{CO}\equiv\int_{\Delta V} S_{CO}\mathrm{d}v>1.0$ Jy km s$^{-1}$ for corresponding rotational molecular transitions CO($J=1-0$)\footnote{This is the most distant line detection with the Yebes 40~m telescope.}, CO($J=2-1$), CO($J=3-2$), and CO($J=5-4$). The binned $\Delta V=500$ km s$^{-1}$ CO-features are detected at the level of $S_{CO}=4.2-23$ mJy, corresponding to signal-to-noise ratios as high as $S/N=48$ in the case of CO($J=3-2$). The transition CO($J=4-3$) is undetected and thus treated as an 3$\sigma$ upper limit estimate\footnote{Close to the edge of the 2 mm window, foreground noise dominates the line signal as this region is heavily contaminated by the atmospheric H$_2$O feature, leading to a non-detection for CO($J=4-3$) at 177.014 GHz.}. We measure a line-averaged redshift of $z_{CO}=1.604543\pm0.00001$.

The CO line profiles match among the transition levels, resembling symmetric double horns with consistent linewidths $\mathrm{FWHM} = 350\pm25$ km s$^{-1}$ (median). This value is on the lower end, however compatible with the FWHM of \citet{Bothwell2013} SMG’s median at $500 \pm 150$ km s$^{-1}$ (MAD).

Strikingly, the line intensities for $J_{upper}\leq3$ are comparable to that of the Cosmic Eyelash at a magnification of $\mu=32$ \citep[lensed by a galaxy cluster at $z=2.3$;][]{Swinbank2010,Danielson2011} while substantial brighter than the strongly lensed \textit{Herschel}-selected SMGs (in principle, galaxy-galaxy lensing) from the \textit{H}-ATLAS catalog compiled by \citet{Yang2017}. Based exclusively on low-$J_{upper}$ lines, SDSSJ1226 is one of the brightest known SMG on the sky. Only \textit{Planck}-selected SMGs host consistently brighter ISMs \citep[see e.g.][]{Harrington2021}. 

As a result of high source magnification, SDSSJ1226 exhibits \emph{ultra-bright} low-$J_{upper}$ CO intensities. However, the trend is reversed for CO($J=5-4$) where we observe reduced brightness, below the average of the \textit{H}-ATLAS comparison sample. Via careful line luminosity ratio measurements, employing the relationship by \citet{Solomon2005} $$ L'_{CO}=3.25\times 10^7\; I_{CO}\;\nu_{obs}^{-2}\;D_L^2\;(1+z)^{-3},$$ in K km s$^{-1}$ pc$^2$ with $\nu_{obs}$ in GHz and $D_L$ in Mpc. To put the intrinsic brightness into context, the line luminosities $L'_{CO(3-2)}$, after correction for magnification, are $(18.2\pm1.5)/\frac{\mu}{9.5}$, $(12.1\pm0.1)/\frac{\mu}{32}$, $(38\pm 8)/\frac{\mu}{6}$, and $9.2\pm2$ $[\times 10^9$ K km s$^{-1}$ pc$^{2}]$ for SDSSJ1226, SMM J2135-0102, \textit{H}-ATLAS SMG mean, and BzK normal galaxies' mean.
More exceptionally, we obtain $r_{5,2}=L'_{CO(5-4)}/L'_{CO(2-1)}=0.11\pm0.02$, the lowest value yet reported during the first four billion years of cosmic time ($z\geq1.5$) \citep{Dannerbauer2009,Daddi2015,Valentino2020}, indicating that individual ISM properties do not necessarily follow a common, universal evolution with redshift \citep[see e.g.][]{Popping2014, Boogaard2020} but can depend strongly on galaxy-wide properties such as e.g. molecular gas fraction. The sub-thermal excitation of typically bright transitions \citep{Bothwell2013, Carilli2013} indicates that the molecular medium of SDSSJ1226 is less dense and/or hot than typically observed for infrared-luminous, $z\gtrsim2$ systems \citep{Papadopoulos2012}. 

\subsection{Additional properties}\label{sec:properties}

\subsubsection{Strong gravitational lensing analysis}\label{sec:lensingmodel}

Since SDSSJ1226 is strongly lensed by a galaxy cluster, the physical interpretation of this galaxy is highly dependent on the magnification factor $\mu$. We employed the public software Lenstool\footnote{\url{https://projets.lam.fr/projects/lenstool/wiki}} \citep{Kneib1993, Jullo2007} to derive the lensing model by utilising the dual pseudo-isothermal elliptical mass distribution (PIEMD) \citep{Limousin2005} for each foreground dark matter halo component. Positions and magnitudes of foreground cluster members were extracted with SExtractor \citep{BertinArnouts1996} from {\it HST}/ACS two-channel maps. We matched visually similar features among the arclets, by colour and morphology. In Fig. \ref{fig:lensing}, the identified image families are shown.
The lensing model predicts a mean magnification in the field observed in the mm-regime of SDSSJ1226-A, in both the northern and southern families combined, of $\mu=9.5\pm 0.7$. It is successful in predicting the approximate locations, 0.2 arcsec rms, of possible counter-images of which three were initially identified (SDSSJ1226-A-north, -A-south, and -B). Tab. \ref{tab:families} provides coordinates and individual magnification factors of the strongly lensed components within the galaxy.
For the central halo, an elliptical mass distribution is predicted. Furthermore, the primary deflector parameter values for SDSSJ1226+2149 are ellipticity $e=0.449$, position angle $P.A. = 36.59^\circ$, core radius $r_\mathrm{core} = 27.64$ kpc, cut radius $r_\mathrm{cut} = 1500$ kpc and normalisation $\sigma_\mathrm{PIEMD} = 865$ km s$^{-1}$. 
The line-of-sight velocity dispersion of the cluster $\sigma_\mathrm{1D}=612$ km s$^{-1}$ is derived from twelve cluster members' spectroscopic redshifts provided in \citet{Bayliss2011}. In order to estimate the size of the cluster, we calculate the radius $R_{200}= 1.21$ Mpc, which approximates the virial radius, and find the virial mass of the cluster to be $M_{200}=1.6\pm0.8\times10^{14}$ M$_\sun$ by employing the relation from \citep{Munari2013}. This cluster mass is in concordance with the mass obtained from our lensing model $M_{lens}(<R_{200})=2.8\times10^{14}$ M$_\sun$. For the fixed redshift of $z=1.6045$, the critical curves are shown in Fig.~\ref{fig:lensing}.

\subsubsection{Active galactic nuclei diagnostic}\label{sec:AGN}
As we decided to expand the NIR/MIR colour criteria to allow a broader diversity of galaxy properties, the possibility of an AGN contribution needs to be excluded since alternatively it could explain the high observed-frame MIR and FIR emission by processes other than star-formation activity. Following the scheme proposed by \citet{Secrest2015} for AllWISE sources, we find observed-frame $[3.4]-[4.6] = 0.3 \pm 0.06$ mag and $[4.6]-[12.0] = -0.2 \pm 0.3$ mag. According to the model tracks relative to the demarcation region of \citet{Mateos2012} these values best match with a aging stellar population at redshift $\sim1.5$, and securely exclude AGN-fractions above $f_\mathrm{AGN}\ge$0.2.

\subsubsection{Panchromatic SED-modelling}\label{sec:SEDfit}
We used the MAGPHYS\footnote{\url{http://www.iap.fr/magphys/index.html}} code \citep{DaCunha2008, DaCunha2015} to compute synthetic spectral energy distributions (SEDs), 
from which we obtained an apparent rest-frame 8-1000 $\mu$m infrared luminosity $\mu L_\mathrm{IR}=1.66^{+0.04}_{-0.04}\times 10^{13}$ L$_\sun$. With the apparent infrared luminosity from the best fitting SED model (reduced $\chi^2=0.63$) we computed the star-formation rate by assuming Salpeter IMF and the conversion factor from \citet{Kennicutt1998}. Accordingly, the apparent star-formation rate of SDSSJ1226 is $\mu\mathrm{SFR}_\mathrm{IR}=2960\pm70$ M$_\sun$ yr$^{-1}$, yielding a magnification corrected star-formation rate of $\mathrm{SFR}_\mathrm{IR}\approx300\times(\mu/9.5)$ M$_\sun$ yr$^{-1}$. MAGPHYS finds a comparatively lower star-formation rate $\mu\mathrm{SFR}=737\pm17$ M$_\sun$ yr$^{-1}$ (accounting for Chabrier-to-Salpeter IMF conversion of $\Upsilon=1.8$), stellar mass $\mu M_{\ast}=1.00^{+0.01}_{-0.02}\times 10^{13}$ M$_\sun$, and mass-weighted age $\mathrm{age}_\mathrm{M}=2.63^{+0.06}_{-0.06}$ Gyr at an average $V$-band attenuation of $A_V=3.61^{+0.1}_{-0.02}$ mag. Although, the stellar mass appears to be enhanced by a factor of $\sim$10 while the stellar populations are $\times$(2-3) older, the posterior values are in broad agreement with the average properties of ALESS $z\leq2.7$ SMGs on the star-forming main-sequence \citep{DaCunha2015,Schreiber2015}.

\section{Discussion} \label{sec:discussion}

\begin{figure}
\epsscale{1.17}
\plotone{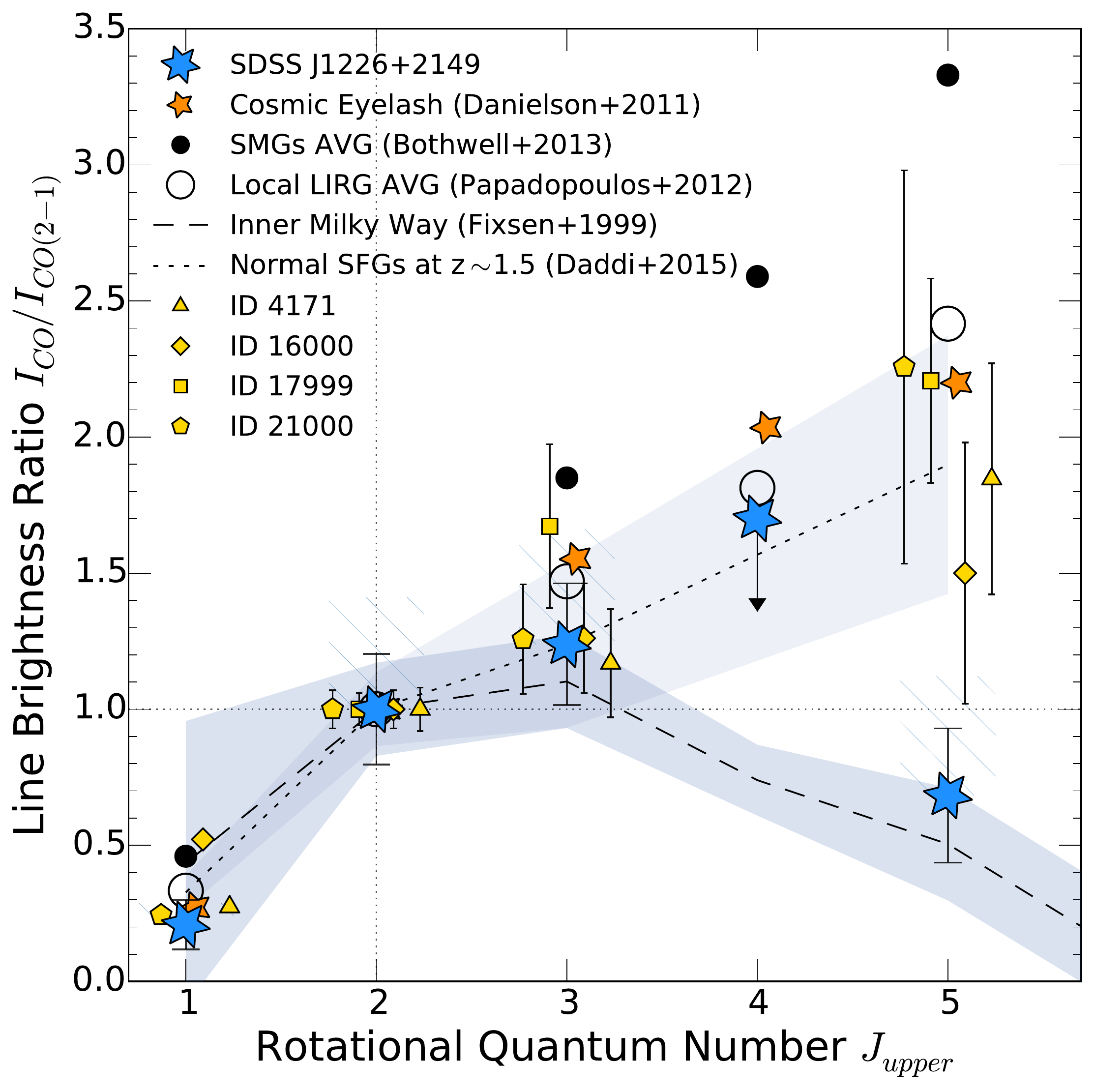}
\caption{CO excitation profile (CO SLED) of SDSSJ1226 (blue stars) normalised to the line intensity of transition CO($J=2-1$). Error bars indicate 3$\sigma$ uncertainty confidence intervals. Line excitation of the inner Milky Way  \citep{Fixsen1999} (dashed; fills for 95\% confidence level) and normal star-forming galaxies at $z$ $\sim$1.5 \citep{Daddi2015} (yellow symbols; fills same as above) with average values thereof are shown as the dotted curve. Median CO excitation for non-lensed SMGs \citep{Bothwell2013} is indicated by  solid circles and local LIRG \citep{Papadopoulos2012} average is shown by open circles. Through transitions CO($J=1-0$) to CO($J=3-2$), SDSSJ1226 and normal SFGs exhibit similar CO line excitation, although, the excitation for transition CO($J=5-4$) of SDSSJ1226 is \emph{significantly lower} than that of normal high-$z$ SFGs, even when considering an under-estimation of the CO line width due to erroneous baseline removal (including $-1\sigma$ shift of baseline level; hatched bars). Intensity for CO($J=4-3$) is an upper-limit measurement.}
\label{fig:COSLED}
\end{figure}

The discovery of this submm-ultrabright source demonstrates the capabilities of our selection method \citep{Diaz-Sanchez2017,Dannerbauer2019} to identify the brigtest SMGs on the sky. The \emph{Cosmic Seahorse} (SDSSJ1226) is a previously unknown FIR-bright galaxy at spectroscopic redshift $z_{CO}=1.60454$ being strongly lensed by a galaxy cluster at $z=0.44$ with magnification factor $\mu=9.5\pm0.7$. \replaced{The source plane image is mapped into at least three sub-images (image A, with image multiplicity of two, and image B) with an optical extend of $\theta_\mathrm{arc}\approx 8.2\arcsec$ for the main source A.}{The galaxy's image is split up into three magnified images -- A-south, A-north with a total extent of $\theta_\mathrm{arc}\approx 8.2\arcsec$, and image B.} Since counter-image B is less \replaced{magnified}{affected by shear}, it might, in future studies, serve as a standard to constrain \replaced{the bias on differential magnification.}{a potential flux bias caused by differential magnification.}

Strong lensing directly affects the obtained IRAM 30~m and Yebes 40~m telescope measurements of the CO lines that are much brighter, without magnification correction, than typically observed for \textit{Herschel}-selected SMGs but comparable in brightness to the CO line flux of the Cosmic Eyelash. However, the CO($J=5-4$) flux and thus warmer/dense gas contribution is significantly below that of the Cosmic Eyelash \citep{Danielson2011} and even below the $J_{upper}=5$ excitation of normal star-forming galaxies at similar redshift \citep{Daddi2015}. To relax our assumption on the shape of the CO SLED, line intensities in Fig. \ref{fig:COSLED} are normalised to $I_{CO(2-1)}$ that is also tracing more diffuse, cirrus-like, and cool gas. Hence, the normalised brightness at high-$J_{upper}$ is used as a proxy diagnostic for the dense and/or warm gas involved in star-formation \citep{Weiss2007}. Typically, 870 $\mu$m-selected SMGs \citep[see e.g.][]{Bothwell2013} show SLEDs with a broad excitation plateau at $J_{upper}\approx5-7$. Similar CO SLEDs are also observed for main sequence star-forming galaxies \citep{Daddi2015, Valentino2020}, although at lower median excitation. Noted by \citet{Popping2014}, the line ratio CO($J=5-4$) over CO($J=2-1$) does scale especially strong with with star-formation activity and molecular gas surface density at all redshifts. Thus a low line luminosity ratio $r_{5,2}$ is commonly interpreted as to broadly trace cold gas conditions \citep[see e.g.][]{Valentino2020} that are ultimately governed by energy and momentum injection from star-formation activity, consequently driving the CO excitation.

Accordingly, CO SLEDs of distant galaxies such as SMGs or normal star-forming galaxies are found to be well explained by a two-component model of the ISM, a cool and warm phase, with intensity contributions depending on the physical properties of the cool gas \citep{Bothwell2013,Daddi2015}. Purely Milky-Way like CO excitation, with very little contribution from dense and warm gas, has never been reported in an early Universe or cosmic noon galaxy before. Although some high-redshift normal SFGs show significantly similar $J_{upper}\leq3$ excitation \citep{Dannerbauer2009,Daddi2015}. Even in consideration of high amounts of low surface-brightness gas -- that might lead to an underestimation of line width and thus could cause issues with background subtraction -- the CO excitation of $J_{upper}=5$ in SDSSJ1226 is by $\sim 3\sigma$ confidence level below the median of  normal SFGs \citep{Daddi2015}. At face value, a CO SLED with $r_{5,2}=0.11\pm0.02$ predominantly resembles single-component excitation that might originate from physical conditions possibly not dissimilar to those observed in the 4 kpc molecular region of the inner Galaxy \citep[see Fig. \ref{fig:COSLED}; ][]{Fixsen1999}. We acknowledge the existence of comparably low $r_{5,2}$-ratios in at least one of the \citet{Valentino2020} reported non-lensed main-sequence galaxies i.e. $r_{5,2}\approx0.13$ in ID 35349 at $z=1.25$. But since none of these sources fulfill both criteria of having (1) sufficient number of high-fidelity, low-$J$ CO flux measurements to confirm \emph{true} Milky Way-like CO-SLEDs at high significance level and (2) can be strictly defined as {\it bona fide} cosmic noon galaxies at $1.5<z<3$, we claim that these sources do not challenge our interpretation of the \emph{Cosmic Seahorse}.

To explain the unusual CO SLED, we consider three possible scenarios. First, measurement errors originating from either calibration, pointing, and/or line width underestimation, might be responsible for the low $J_{upper}=5$ excitation. However, only an unfortunate combination of these observational effects together could  produce the magnitude of the effect in the data. 

Second, a short-lived phenomenon -- like a tidal bridge, commonly seen in LIRGs -- might indeed produce similarly low excitation mainly due to a local decrease of high-density gas fraction \citep{Zhu2003,Weiss2007}. Together with a possible magnification bias, this spatially confined component might then be preferentially amplified. We deem this scenario as less likely, since our lensing map does not predict strong differential magnification beyond $\gtrsim10 \%$.

Third, a previously unseen mechanism of early Universe star-formation, acting on high mass, low $L_\mathrm{IR}$ systems, might be present. The last scenario is the most attractive, as the number counts of SMGs are intrinsically steep \citep[see e.g.][]{Negrello2010} and although IR-luminous systems, ULIRGs and HyLIRGs, dominate star-formation activity beyond $z\gtrsim1$ \citep{LeFloch2005} still a large fraction of cosmic star-formation rate density occurs in $L_\mathrm{IR}\lesssim10^{12}$ L$_\sun$ systems \citep{Rodighiero2011}. 
From a theoretical perspective, a correlation between CO excitation and star-formation surface density is expected. According to \citet{Narayanan2014} model predictions\footnote{\url{https://sites.google.com/a/ucsc.edu/krumholz/codes/co-sled}}, with relatively mild star-formation surface density $\Sigma_{\mathrm{SFR}}=0.46\pm0.02$ M$_\sun$ yr$^{-1}$ kpc$^{-2}$ measured within the area of the main optical arc, we find the line luminosity ratio $r_{5,2}(\mathrm{\Sigma_\mathrm{SFR}})=0.25\pm0.01$. The model CO SLED is in concordance with low-excitation gas and broadly with the observed value $r_{5,2}=0.11\pm0.02$ at reduced $\chi_r^2=1.21$ for the overall SLED.

Following this line of evidence, we conclude that the Cosmic Seahorse might most likely intrinsically belong to an under-explored population of dusty, low  $L_\mathrm{IR}$ systems with high gas masses -- in disguise of classical ULIRGs -- that form stars in extended, clumpy discs at low efficiency. We have no strong argument to support a major merger scenario, like broad line widths, but can also not refute it. Additional measurements at high-$J$ CO transitions, resolved interferometry of the continuum, and better constrains on the source magnification are indispensable to further eliminate one of the proposed hypotheses.


Given the remarkable low $r_{5,2}$ line ratio, and disregarding magnification bias, we postulate that the \emph{Cosmic Seahorse} (SDSSJ1226) \replaced{could host}{hosts} \textit{Milky Way-like} cold gas conditions -- similar to that seen in the inner Galactic region -- but in the distant Universe. Contrary to the more typical high-excitation regime of cold gas in SMGs, this novel source with overall star-forming main-sequence characteristics shows unique interstellar medium properties for a strongly lensed, ultra-bright galaxy at $S_{500\mu m}>100$ mJy. Providing further circumstantial evidence to this hypothesis, we find that the locus of intrinsic stellar mass over star-formation rate, including all discussed uncertainties (see Sec. \ref{sec:properties}), falls within $^{+1}_{-2}\times\sigma$ of that of main-sequence galaxies \citep{Schreiber2015} at the same cosmic epoch.

Moreover, in virtue of its bright CO emission lines, SDSSJ1226 could also serve as a new reference source for extended, low efficiency star-formation at high-redshift. Follow-up observations should be able to spatially resolve the main giant arc -- in analogy to resolved studies of SMM J2135-0102 \citep{Swinbank2011,Ivison2020} -- into individual giant molecular clouds, further enabling insight into a massive galaxy at a truly complementary track to local ULIRG-like galaxy evolution. At declination $\delta\simeq+22^\circ$, both ALMA and IRAM NOEMA interferometers would be capable of expanding line observations beyond CO($J=5-4$) at high spatial resolution to ultimately verify the unusual CO excitation profile among the \emph{Cosmic Seahorse'} molecular clouds. The discovery of low gas excitation in an otherwise typical dusty star-forming galaxy emphasizes the inferred diversity of yet unexplored pathways in early galaxy assembly. 

\acknowledgments

We thank the anonymous referee for providing constructive comments that helped to improve the quality of our work.
This work is based on observations carried out under project number 086-18 with the IRAM 30m telescope. IRAM is supported by INSU/CNRS (France), MPG (Germany) and IGN (Spain). We acknowledge IRAM AoD Wanju Kim for her support. Based on observations with the 40-m radio telescope of the National Geographic Institute of Spain (IGN) at Yebes Observatory (project nr. 20B013). Yebes Observatory thanks the ERC for funding support under grant ERC-2013-Syg-610256-NANOCOSMOS. N.S. acknowledges the support from the University of Vienna through the "short-term grants abroad" programme (KWA). H.D. acknowledges financial support from the Spanish Ministry of Science, Innovation and Universities (MICIU) under the 2014 Ramon y Cajal program RYC-2014-15686 and under the AYA2017-84061-P, co-financed by FEDER (European Regional, Development Funds), and in addition, from the Agencia Estatal de Investigacion del Ministerio de Ciencia e Innovaci\'{o}n (AEI-MCINN) under grant (La evoluci\'{o}n de los c\'{u}umulos de galaxias desde el amanecer hasta el mediod\'{i}a c\'{o}smico) with reference (PID2019-105776GB-I00/DOI:10.13039/501100011033). A.D.S., S.I.G. and R.L.L. acknowledge financial support by project “Participation in the NISP instrument and preparation for the scientific exploitation of Euclid”, PID2019-110614GB-C22/AEI/10.13039/501100011033, \& PID2019-110614GB-C21 financed by the “Agencia Estatal de Investigación” (AEI-MCINN). This publication makes use of data products from the \textit{Wide-field Infrared Survey Explorer}, which is a joint project of the University of California, Los Angeles, and the Jet Propulsion Laboratory/California Institute of Technology, funded by the National Aeronautics and Space Administration. We acknowledge the use of GILDAS software (\url{http://www.iram.fr/IRAMFR/GILDAS}).  Herschel is an ESA space observatory with science instruments provided by European-led Principal Investigator consortia and with important participation from NASA; we used data from proposal ID GT1\_dlutz\_4: observations 1342200240 (Herschel/SPIRE) and 1342221962 (Herschel/PACS). Based on observations made with the NASA/ESA {\it Hubble} Space Telescope, obtained from the Data Archive at the Space Telescope Science Institute, which is operated by the Association of Universities for Research in Astronomy, Inc., under NASA contract NAS 5-26555. These observations are associated with program 12368. 

\bibliographystyle{aasjournal}

\appendix
\section{Lensing map}

\begin{figure}[h!]
\epsscale{1.17}
\plotone{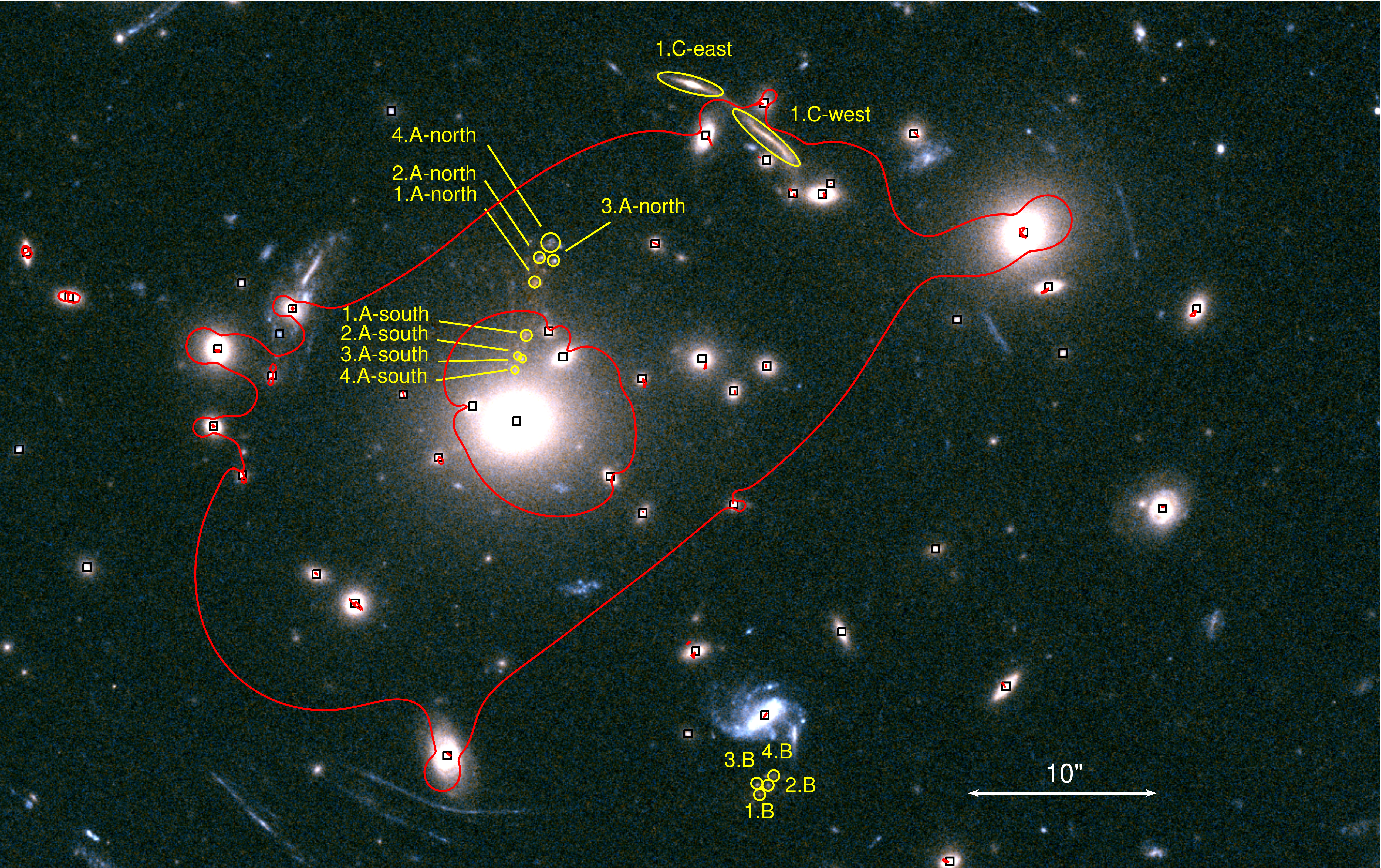}
\caption{Two-channel \textit{HST}/ACS map showing filter bands F555W (cyan) and F814W (orange) with over-plotted lensing analysis constrains. Foreground galaxy cluster members at $z\approx0.435$ are indicated as black squares. The rest-frame UV-bright clumps that are used to model the gravitational lensing effect in the three images of the galaxy SDSSJ1226 (A-north, A-south, B) are marked in yellow and there are four families (1, 2, 3, and 4). Galaxy SDSSJ1226-C is the same redshift as our target, yet a distinct system, and also lensed into two images, C-east and C-west. Shown in red are the modelled critical curves of the foreground galaxy cluster potential, at the same redshift as the Cosmic Seahorse and source SDSSJ1226-C.
}
\label{fig:lensing}
\end{figure}

\begin{deluxetable}{lccc}
\tablenum{2}
\tabletypesize{\footnotesize}
\tablecaption{List of constrains for the lensing model by matching image families among the arclets. Coordinates are for epoch J2000.\label{tab:families}}
\tablehead{\colhead{ID $^\mathrm{a}$} & \colhead{R.A. [$\degr$]} & \colhead{Dec. [$\degr$]} & \colhead{$\mu$ $^\mathrm{b}$} }
\decimalcolnumbers
\startdata
1.A-north & 186.71274 & 21.83322  &  8.1$\pm$1.1 \\
1.A-south & 186.71290 & 21.83244  &  4.4$\pm$0.8 \\
1.B       & 186.70915 & 21.82564  &  2.2$\pm$0.3 \\
\hline
2.A-north & 186.71267  &  21.83357 &   7.3$\pm$0.7 \\
2.A-south & 186.71302  &  21.83212  &  1.9$\pm$0.4 \\
2.B       & 186.70902 &   21.82578 &   2.3$\pm$0.3 \\
\hline
3.A-north & 186.71244  &  21.83355  &  6.0$\pm$0.6 \\
3.A-south & 186.71296 &   21.83210  &  1.7$\pm$0.4 \\
3.B       & 186.70919 &   21.82579  &  2.3$\pm$0.3 \\
\hline
4.A-north & 186.71251 &   21.83381  &  7.2$\pm$0.6 \\
4.A-south & 186.71306 &   21.83192 &   1.2$\pm$0.3 \\
4.B       & 186.70894  &  21.82592 &   2.3$\pm$0.3 \\
\hline\\
\hline
1.C-east & 186.71021  &  21.83615  & 12$\pm$3 \\
1.C-west & 186.70894  &  21.83530  & 17$\pm$5
\enddata
\tablecomments{
$^\mathrm{a}$All entries start formally with 'SDSSJ1226-'. The digit refers to the ID of the family of images coming from the same region within the source galaxy. The letter and positional argument indicates the ID of the image seen in the image plane. $^\mathrm{b}$Magnification factor.
}
\end{deluxetable}




\end{document}